\documentclass[sigconf,nonacm]{acmart}

\AtBeginDocument{\providecommand\BibTeX{{\normalfont B\kern-0.5pt{\scshape i\kern-0.25pt b}\kern-0.8pt\TeX}}}
\usepackage{totpages}
\usepackage{float}
\usepackage{tcolorbox}
\usepackage{fontawesome5} 
\usepackage{enumitem}     
\usepackage{needspace,etoolbox}
\usepackage{balance}

\renewcommand\footnotetextcopyrightpermission[1]{}
\title{Integrating Large Language Models in Software Engineering Education: A Pilot Study through GitHub Repositories Mining}
\renewcommand{\shorttitle}{Integrating LLMs in SE Education}

\author{Maryam Khan}
\email{maryam.khan@student.lut.fi}
\affiliation{%
  \institution{Software Engineering Department, Lappeenranta-Lahti University of Technology}
  \city{Lappeenranta}
  \country{Finland}
}

\author{Muhammad Azeem Akbar}
\email{azeem.akbar@lut.fi}
\affiliation{%
  \institution{Software Engineering Department, Lappeenranta-Lahti University of Technology}
  \city{Lappeenranta}
  \country{Finland}
}

\author{Jussi Kasurinen}
\email{Jussi.Kasurinen@lut.fi}
\affiliation{%
  \institution{Software Engineering Department, Lappeenranta-Lahti University of Technology}
  \city{Lappeenranta}
  \country{Finland}
}

\renewcommand{\shortauthors}{Khan et al.}

\begin{document}

\begin{abstract}
\textbf{Context:} Large Language Models (LLMs) such as ChatGPT are increasingly adopted in software engineering (SE) education, offering both opportunities and challenges. Their adoption requires systematic investigation to ensure responsible integration into curricula.  \textbf{Objective:} This doctoral research aims to develop a validated framework for integrating LLMs into SE education through a multi-phase process, including taxonomies development, empirical investigation, and case studies. This paper presents the first empirical step. \textbf{Method:} We conducted a pilot repository mining study of 400 GitHub projects, analyzing README files and issues discussions to identify the presence of motivator and demotivator previously synthesized in our literature review \cite{khan2024llmsreview} study. \textbf{Results:} Motivators such as engagement and motivation (227 hits), software engineering process understanding (133 hits), and programming assistance \& debugging support (97 hits) were strongly represented. Demotivators, including plagiarism \& IP concerns (385 hits), security, privacy \& data integrity (87 hits), and over-reliance on AI in learning (39 hits), also appeared prominently. In contrast, demotivators such as challenges in evaluating learning outcomes and difficulty in curriculum redesign recorded no hits across the repositories. \textbf{Conclusion:} The study provides early empirical validation of motivators/demotivators taxonomies with respect to their themes, highlights research–practice gaps, and lays the foundation for developing a comprehensive framework to guide the responsible adoption of LLMs in SE education.  
\end{abstract}



\maketitle

\section{Introduction} \label{sec:Introduction}

Large Language Models (LLMs) based tools are rapidly transforming the landscape of software engineering (SE) education. Their ability to generate code, provide tailored explanations, and support problem-solving has attracted widespread attention from students, educators, and researchers \cite{imai2022github}. These tools promise to lower barriers in programming education, foster adaptive learning, and increase student engagement. However, their integration is not without pitfalls: risks such as data leakage, plagiarism, ethical concerns, over-reliance on automation, and reproducibility issues have been highlighted as critical challenges both in SE education and research domains \cite{akbar2023ethical, sallou2024breaking}.

Our recent literature review synthesized these opportunities and risks into a structured taxonomies of motivators (success factors) and demotivators (challenges) for LLM use in SE education \cite{khan2024llmsreview}. The taxonomies consist of \textit{four high-level motivator themes with nine sub-themes} and \textit{four high-level demotivator themes with ten sub-themes}. The motivator sub-themes include, for example, \emph{Programming Assistance and Debugging Support}, \emph{Personalized and Adaptive Learning}, \emph{Conceptual Understanding and Problem-Solving}, and \emph{Engagement and Motivation}. The demotivator sub-themes include, among others, \emph{Plagiarism and Intellectual Property Concerns}, \emph{Over-Reliance on AI in Learning}, \emph{Bias and Hallucination in Outputs}, and \emph{Security, Privacy, and Data Integrity Issues}. The texonomies not only provided a comprehensive conceptual overview but also served as the foundation for our empirical study, which operationalized the sub-themes as units of analysis. 

Building on this foundation, this research project aims to deliver a validated framework for integrating LLMs into SE education through four interconnected phases. The framework will be developed in four phases: (i) constructing an automated decision making taxonomies of motivators and demotivators by combining literature insights with empirical evidence; (ii) refining this taxonomies through surveys and interviews with students, educators, and management level experts; (iii) deriving design principles for curriculum and tool integration; and (iv) validating the framework through case studies. As a first empirical step, this paper presents a pilot repository mining study of 400 GitHub projects that incorporate LLMs into educational contexts. By analyzing README files and issue discussions, we compare practical experiences with motivators and demotivators themes identified in the literature review study \cite{khan2024llmsreview}, highlighting both areas of alignment and divergence.

\section{Background and Motivation}

Recent advances in LLMs have prompted significant interest in their role within SE education. Early empirical studies show that LLMs can enhance programming assistance, debugging, and problem-solving, thereby lowering entry barriers for novice learners and supporting personalized learning pathways \cite{imai2022github}. Similarly, LLMs have been found to foster engagement and provide adaptive learning support, positioning them as potentially transformative educational technologies \cite{akbar2023ethical, khan2024llmsreview}.  

Despite these benefits, concerns remain. Scholars have highlighted risks of plagiarism and academic dishonesty \cite{akbar2023ethical}, over-reliance on AI tools that may erode critical thinking skills \cite{khan2024llmsreview}, ethical and privacy issues in data handling \cite{akbar2023ethical}, and hallucinations or unreliable outputs. These challenges highlight the dual nature of LLM adoption in SE education, where potential motivators (success factors) coexist with demotivators (barriers).  

To synthesize these insights, our recent literature review developed taxonomies that organize motivators and demotivators into high-level themes and sub-themes. Although the taxonomies synthesized from the review provide a comprehensive conceptual overview, most of the primary studies in the literature are based on controlled experiments or hypothetical scenarios, with limited evidence from real-world educational adoption.

To address this limitation, we turned to GitHub\footnote{https://github.com/}, a platform widely used by educators, researchers, and developers to share projects and tools. Mining GitHub repositories provides a valuable perspective into how LLMs are practically explored in educational contexts, capturing community-driven experimentation, documentation, and open discussions of opportunities and challenges. By mapping repository evidence onto the motivator and demotivator themes, this study not only validates several insights from academic literature but also reveals new, practice-oriented dimensions of adoption that are often overlooked in theoretical discourse. 
\section{Methodology} \label{Sec:methodology}
\begin{figure*}[t!]
  \centering
  \includegraphics[width=\linewidth]{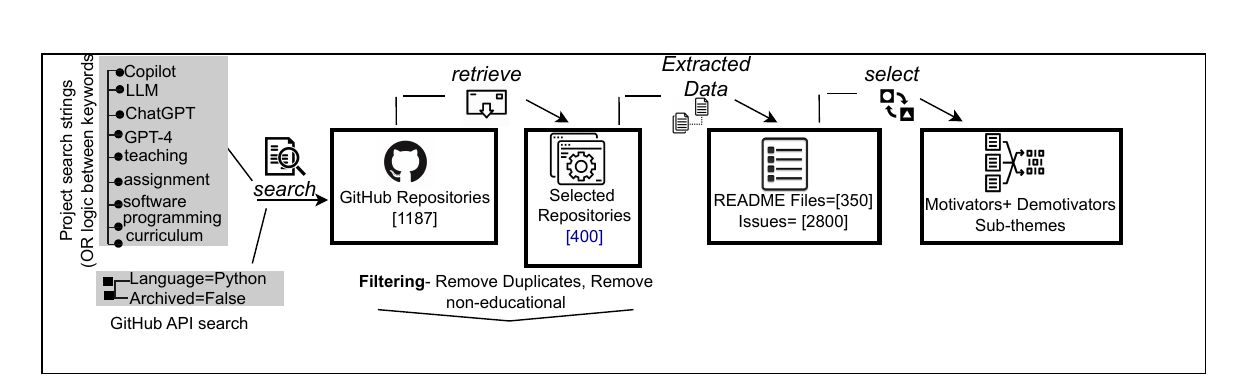}
  \caption{Overview of the Methodology}
  \label{fig:repo-mining-process}
\end{figure*}

This research follows a multiphase design to develop and validate a framework for integrating LLMs into SE education: as discussed in Section \ref{sec:Introduction}. This paper addresses phase (i) through a pilot repository mining study, and the methodological process is illustrated in Figure~\ref{fig:repo-mining-process}. A key design choice was to operationalize the taxonomies of motivators and demotivators at the sub-theme level. Sub-themes capture concrete manifestations (e.g., \textit{debugging support}, \textit{plagiarism and IP concerns}, \textit{bias and hallucinations}) that align closely with repository artifacts such as README files and issue discussions, making them better suited for empirical detection than broader conceptual themes. This fine-grained operationalization also helps mitigate known validity threats in GitHub mining, where coarse or ambiguous categories risk misclassification and bias \cite{kalliamvakou2014promises}. In doing so, we were able to test whether detailed factors identified in the literature are also reflected in real-world educational projects.

Following established repository mining practices \cite{kalliamvakou2014promises}, we constructed GitHub queries combining LLM keywords included ``ChatGPT'', ``Copilot'', ``LLM'', ``GPT-4'', ``assignment'', ``software programming'', and ``curriculum''. To ensure consistency, we restricted results to Python repositories. Using the GitHub API, we collected metadata, README files, and issue discussions aligned with the taxonomy. From 1187 repositories, filtering removed duplicates, inactive, and non-educational projects, yielding 400 repositories (350 README files and 2800 issue discussions, max. 8 per repository). The cap prioritized breadth over depth, consistent with prior work \cite{munaiah2017curating}. Issues were selected via keyword-guided queries mapped to sub-themes (e.g., grading'', feedback'', plagiarism'', hallucination''), while irrelevant bug reports or installation errors were excluded. The replication package is provided in \cite{khan2025llm}, including the dataset of the GitHub projects extracted, the data extraction, and the analysis scripts. Analysis followed the taxonomies from our earlier work \cite{khan2024llmsreview}. Each repository was coded as a hit (at least one sub-theme keyword instance) or non-hit. To test representation of motivator/demotivator sub-themes, we applied a Chi-square goodness-of-fit test, assessing whether observed distributions deviated from equal expectations at the 95\% confidence level.

\section{Results and Discussion}
Mining of 400 GitHub projects provided empirical evidence that both supports and extends our earlier literature review \cite{khan2024llmsreview}. Table~\ref{tab:chi-square} summarizes the chi–square goodness‑of‑fit tests, where we examined whether observed hit/non-hit distributions for motivator and demotivator sub-themes deviated from an equal-probability baseline ($H_{0}$: hits = non-hits) toward adoption-consistent patterns ($H_{1}$).

\begin{table}[htbp]
\centering
\caption{Chi-square goodness-of-fit results for motivator and demotivator sub-themes (baseline: hits = non-hits)}
\label{tab:chi-square}
\resizebox{\linewidth}{!}{%
\begin{tabular}{|l|c|c|c|c|c|}
\hline
\textbf{Theme} & \textbf{With Hits} & \textbf{Without Hits} & $\chi^2$ & \textbf{p-value} & \textbf{Category} \\
\hline
Software Engineering Process Understanding & 133 & 267 & 44.89  & $<0.001$ & Motivator \\
\hline
Personalized and Adaptive Learning         & 49  & 351 & 228.01 & $<0.001$ & Motivator \\
\hline
Engagement and Motivation                  & 227 & 173 & 7.29   & 0.007     & Motivator \\
\hline
Conceptual Understanding \& Problem Solving & 69  & 331 & 171.61 & $<0.001$ & Motivator \\
\hline
Programming Assistance \& Debugging Support & 97  & 303 & 106.09 & $<0.001$ & Motivator \\
\hline
Automated Assessment \& Grading            & 2   & 398 & 392.04 & $<0.001$ & Motivator \\
\hline
Formative Feedback \& Learning Support     & 16  & 384 & 338.56 & $<0.001$ & Motivator \\
\hline
AI as a Learning Partner                   & 4   & 396 & 384.16 & $<0.001$ & Motivator \\
\hline
Project-Based \& Inquiry-Based Learning    & 1   & 399 & 396.01 & $<0.001$ & Motivator \\
\hline
Plagiarism \& IP Concerns                  & 385 & 15  & 342.25 & $<0.001$ & Demotivator \\
\hline
Security, Privacy \& Data Integrity        & 87  & 313 & 127.69 & $<0.001$ & Demotivator \\
\hline
Over-Reliance on AI in Learning            & 39  & 361 & 259.21 & $<0.001$ & Demotivator \\
\hline
Computational \& Resource Costs            & 37  & 363 & 265.69 & $<0.001$ & Demotivator \\
\hline
Bias \& Hallucination in Outputs           & 15  & 385 & 342.25 & $<0.001$ & Demotivator \\
\hline
Limitations in Understanding \& Context    & 14  & 386 & 345.96 & $<0.001$ & Demotivator \\
\hline
Reduced Critical Thinking                  & 11  & 389 & 357.21 & $<0.001$ & Demotivator \\
\hline
Ethical Concerns in AI Learning            & 9   & 391 & 364.81 & $<0.001$ & Demotivator \\
\hline
Challenges in Evaluating Learning Outcomes & 0   & 400 & 400.00 & $<0.001$ & Demotivator \\
\hline
Difficulty in Curriculum Redesign          & 0   & 400 & 400.00 & $<0.001$ & Demotivator \\
\hline
\end{tabular}}
\end{table}
\vspace{-0.5em}

\begin{itemize}
\item \textbf{Interpreting significance vs. prevalence}:We emphasize that the test evaluates deviation from an \emph{equal} hit/non-hit baseline rather than raw prevalence. Accordingly, some sub-themes with relatively \emph{low} hit counts still yield very small $p$-values because they are markedly \emph{under}-represented relative to $H_{0}$ (e.g., \textit{Automated Assessment \& Grading}: 2/400; \textit{AI as a Learning Partner}: 4/400; \textit{Project-Based \& Inquiry-Based Learning}: 1/400). With $n=400$, even small departures from the null can become statistically significant; hence, statistical significance should not be conflated with practical maturity \cite{agresti2007introduction}.
\item  \textbf{Motivators}: \textit{Engagement and Motivation} (227/400; $\chi^{2}=7.29$, $p=0.007$) is the only motivator observed in a majority of repositories, indicating consistent salience in public project artifacts. \textit{Software Engineering Process Understanding} (133/400; $\chi^{2}=44.89$, $p<0.001$), \textit{Programming Assistance \& Debugging Support} (97/400; $\chi^{2}=106.09$, $p<0.001$), and \textit{Conceptual Understanding \& Problem Solving} (69/400; $\chi^{2}=171.61$, $p<0.001$) also deviate from the equal baseline, though their absolute counts are lower. Conversely, \textit{Automated Assessment \& Grading} (2/400), \textit{AI as a Learning Partner} (4/400), \textit{Formative Feedback \& Learning Support} (16/400), and \textit{Project-Based \& Inquiry-Based Learning} (1/400) are \emph{rare} in the corpus yet statistically significant owing to strong under-representation relative to $H_{0}$.
\item \textbf{Demotivators}: \textit{Plagiarism \& IP Concerns} (385/400; $\chi^{2}=342.25$, $p<0.001$) dominates repository discussions, underscoring academic integrity as a principal concern in practice. \textit{Security, Privacy \& Data Integrity} (87/400; $\chi^{2}=127.69$, $p<0.001$) and \textit{Over-Reliance on AI in Learning} (39/400; $\chi^{2}=259.21$, $p<0.001$) are also consistently present. Other demotivators (e.g., \textit{Bias \& Hallucination in Outputs} 15/400; \textit{Reduced Critical Thinking} 11/400; \textit{Ethical Concerns in AI Learning} 9/400) appear less frequently but still deviate from the equal baseline. Notably, \textit{Challenges in Evaluating Learning Outcomes} and \textit{Difficulty in Curriculum Redesign} recorded \emph{zero} hits (each $\chi^{2}=400$, $p<0.001$), indicating complete under-representation relative to $H_{0}$. Substantively, their absence in public repositories suggests these curriculum-level issues may surface later in institutional adoption or may be under-reported in open-source channels.
\item \textbf{Multiple testing and robustness}: 
Because we tested 19 sub-themes, some small $p$-values could occur by chance. We therefore controlled the false discovery rate using the Benjamini–Hochberg procedure (FDR $=0.05$) \cite{benjamini1995controlling}. After correction, all results we mark as significant remained significant (the largest raw $p$-value was 0.007 for \emph{Engagement and Motivation}). To separate statistical deviation from practical importance, we report raw hit counts with $p$-values and caution that significance reflects deviation from the equal-baseline null, not the prevalence or magnitude of a theme. 
\end{itemize}

\section{Roadmap}
The results of this pilot empirical study, together with the literature review findings \cite{khan2024llmsreview}, demonstrate the feasibility of advancing the project through its subsequent phases.  

In the next phase, repository mining will be extended beyond GitHub to include other platforms such as GitLab, Bitbucket, SourceForge, and Kaggle. This broader scope will provide a richer empirical foundation for constructing automated taxonomies of motivators and demotivators by capturing diverse practices, communities, and governance structures.  Building on this repository evidence, additional data will be collected through surveys and semi-structured interviews with students, educators, and management-level stakeholders. These complementary sources will enable the taxonomies to be further refined and automated using evolutionary approaches to decision-making and classification \cite{deb2011multi}.  

The refined and automated taxonomies will then serve as the basis for deriving design principles for curriculum and tool integration, ensuring actionable guidance for higher education contexts. In the long term, this phased program of work will deliver a comprehensive and empirically validated framework to support the responsible integration of LLMs in SE education. By combining large-scale repository evidence with expert validation, algorithmic refinement, and case-based testing, the project will provide actionable insights that bridge the gap between research and practice in AI-driven education.
\begin{tcolorbox}[colback=black!5, colframe=black!20, width=\columnwidth, arc=1mm, auto outer arc, boxrule=1.0pt, left=3mm] \faHandPointRight\hspace{0.5em}\textbf{Key Findings:} \begin{itemize}[leftmargin=2em, nosep] \item Alignment with the literature for prominent motivators (\emph{engagement \& motivation}, \emph{software engineering process understanding}, \emph{programming assistance \& debugging support}) and demotivators (\emph{plagiarism \& IP concerns}, \emph{security, privacy \& data integrity}). \item Several motivators are \emph{rare yet significant} due to under-representation relative to the equal-baseline null (e.g., \emph{project-based \& inquiry-based learning}, \emph{automated assessment \& grading}, \emph{AI as a learning partner}). \item Curriculum-level demotivators (\emph{challenges in evaluating learning outcomes}, \emph{difficulty in curriculum redesign}) are absent in the corpus (0/400), pointing to timing/reporting gaps between theory and public practice. \end{itemize} \end{tcolorbox}

    \textbf{Declaration of generative AI and AI-assisted technologies in the writing process.} \\
During the preparation of this work, the author(s) used ChatGPT to assist with grammatical and other writing-related issues. After using this tool, the author(s) reviewed and revised the content as needed and take full responsibility for the final publication.

\textbf{Declaration of Competing Interest.} 

The authors declare that they have no known competing financial interests or personal relationships that could have appeared to influence the work reported in this document.

\
\balance
\bibliographystyle{ACM-Reference-Format}
\bibliography{main}

\begin{thebibliography}{99}
\bibitem{imai2022github} Imai, Saki (2022). Is github copilot a substitute for human pair-programming? an empirical study. Proceedings of the ACM/IEEE 44th International Conference on Software Engineering: Companion Proceedings. 319--321
\bibitem{akbar2023ethical} Akbar, Muhammad Azeem, Khan, Arif Ali, Liang, Peng (2023). Ethical aspects of ChatGPT in software engineering research. IEEE Transactions on Artificial Intelligence. 
\bibitem{sallou2024breaking} Sallou, June, Durieux, Thomas, Panichella, Annibale (2024). Breaking the silence: the threats of using llms in software engineering. Proceedings of the 2024 ACM/IEEE 44th International Conference on Software Engineering: New Ideas and Emerging Results. 102--106
\bibitem{khan2024llmsreview} Khan, Maryam, Akbar, Muhammad Azeem, Kasurinen, Jussi (2025). Integrating LLMs in Software Engineering Education: Motivators, Demotivators, and a Roadmap Towards a Framework for Finnish Higher Education Institutes. arXiv preprint arXiv:2503.22238. 
\bibitem{kalliamvakou2014promises} Kalliamvakou, Eirini, Gousios, Georgios, Blincoe, Kelly, Singer, Leif, German, Daniel M, Damian, Daniela (2014). The promises and perils of mining github. Proceedings of the 11th working conference on mining software repositories. 92--101
\bibitem{munaiah2017curating} Munaiah, Nuthan, Kroh, Steven, Cabrey, Craig, Nagappan, Meiyappan (2017). Curating GitHub for engineered software projects. Empirical Software Engineering. 3219--3253
\bibitem{khan2025llm} Khan, Maryam (2025). Integrating Large Language Models in Software Engineering Education: A Pilot Study through GitHub Repository Mining. . 
\bibitem{agresti2007introduction} Agresti, Alan (2007). An Introduction to Categorical Data Analysis. Wiley-Interscience. 
\bibitem{benjamini1995controlling} Benjamini, Yoav, Hochberg, Yosef (1995). Controlling the false discovery rate: a practical and powerful approach to multiple testing. Journal of the Royal Statistical Society: Series B (Methodological). 289--300
\bibitem{deb2011multi} Deb, Kalyanmoy (2011). Multi-objective optimisation using evolutionary algorithms: an introduction. Multi-objective evolutionary optimisation for product design and manufacturing. 3--34
\end{thebibliography}

\end{document}